\documentclass[]{spie}  

\usepackage{amsmath,amsfonts,amssymb}
\usepackage{graphicx}
\usepackage[colorlinks=true, allcolors=blue]{hyperref}
\usepackage{color, soul}

\title{Automated testing of optical fibres: towards the design of the Maunakea Spectroscopic Explorer Fibre Transmission System}

\author[a\footnote{Email: montys@uvic.ca}]{Stephanie Monty}
\author[a]{Farbod Jahandar}
\author[a]{Jooyoung Lee}
\author[a]{Kim A. Venn}
\author[a]{Colin Bradley}
\author[b]{Darren Erickson}
\author[b]{David Crampton}
\author[a]{Victor Nicolov}
\author[a]{Collin L. Kielty}
\author[c]{Celine Mazoukh}
\author[d]{Patrick Hall}
\author[a, c]{The FiTS Team}

\affil[a]{University of Victoria, 3800 Finnerty Rd, Victoria, BC, V8P 5C2, Canada}
\affil[b]{National Research Council Herzberg, 5071 W. Saanich Rd., Victoria, BC, V9E 2E7, Canada}
\affil[c]{FiberTech Optica Inc., 330 Gage Avenue Unit 1, Kitchener, ON, N2M 5C6, Canada}
\affil[d]{York University, 4700 Keele St, Toronto, ON, M3J 1P3, Canada}

\begin{document} 
\maketitle
\begin{abstract}
We present the results of an automated fibre optic test bench constructed at the University of Victoria as part of the Maunakea Spectroscopic Explorer (MSE) Fibre Transmission System (FiTS). In preparation for MSE-FiTS, we have begun characterizing the focal ratio degradation (FRD) of candidate multi-mode fibres with the ultimate goal of testing all $\sim4000$ MSE fibres. To achieve this, we have built an optical bench to perform an automated version of the collimated beam test. Herein we present the design of the bench and discuss the automation of components by introducing the Big FiTS Fibre Wrapper (Big FFW), our open-source automation software. We conclude with the results of tests performed using the Big FFW on a sample of candidate fibre, comparing the Big FFW results against those found using manual methods. Our results suggest that the candidate MSE fibre meets the science requirement of $<5\%$ FRD at $f/2$ and find less than $1\%$ disagreement between both measurement methods.
\end{abstract}


\section{INTRODUCTION}
\label{sec:intro}  
The Maunakea Spectroscopic Explorer (MSE) is a proposed upgrade of the current 3.6m Canada-France-Hawaii Telescope into a 11.25m aperture,  1.5 square degree field of view telescope, fully dedicated to performing multi-object spectroscopy \cite{mccon16,saun16,zhang16}. MSE will provide a spectral resolution performance of $R\sim 2500-40000$ across the full wavelength range of $0.36-1.8\mu$m. To accomplish this, the MSE Fibre Transmission System (FiTS) will make use of more than 4000 multi-mode optical fibres with the intent of near identical light delivery among fibres to within 1\%. Through a specially designed fibre positioning system and fibre bundling assembly, FiTS will transmit the light from the prime focus $\sim 50$m to the low-resolution spectrographs located in the telescope pier and $\sim 35$m to the high-resolution spectrographs located on the Nasmyth platforms (see paper number 10702-284 of the same proceedings for more details\cite{venn18}). As the performance of FiTS is subject to both fibre characteristics and telescope dynamics, both need to be well understood prior to the construction of MSE. This requires the successful verification of fibre performance through measurements of individual fibres and fibre bundles, such as; fibre focal ratio degradation (FRD), throughput, general wavelength dependency, fibre cross-talk and any effects associated with fibre capping/splicing procedures. Additionally, the simulated effects of on-sky telescope dynamics on fibres must be tested. As part of the MSE FiTS project, we have constructed an automated fibre optic test bench to perform in-direct FRD measurements of MSE candidate fibres in a fast, efficient and consistent manner. This as an important facility not just in the context of MSE but also in the broader context of future fibre-fed instruments and multi-fibre positioning systems like TMT's WFOS\cite{paz06} and E-ELT's MOSAIC\cite{kelz15} instruments and GMT's MANIFEST\cite{good12} fibre positioner. Here we present the details of our optical bench including; the optical components involved, automation hardware, and the software created to facilitate both measurements and analysis of the fibre FRD.

\section{\label{sec:mantest} MANUAL TESTING OF FIBRE FRD}
Fibre FRD (described in Section \ref{sec:theory}) is tested using one of two methods, the ``formed beam'' test or the collimated (``ring'') test corresponding to a direct and an in-direct measurement of FRD respectively. In the past, optical fibres used in astronomical applications, like the Gemini Remote Access CFHT ESPaDOnS\footnote{Echelle SpectroPolarimetric Device for the Observation of Stars at CFHT\cite{don13}} Spectrograph (GRACES) 270m long fibre (Pazder et al., 2014)\cite{paz14}, were characterized using both methods, adapting the test methodology described by Carrasco\cite{car94}~, Barden \& Ramsey\cite{ram88} and Bershady\cite{bersh04}~. Both tests require the assembly of a dedicated optical bench, with the full beam test requiring careful consideration of bench geometry in order to measure the output f/\# of the fibre directly. Though the ring test requires a  simpler set-up than the full beam test, careful consideration of camera and image characteristics is required to ensure a true proxy to FRD is found.

\subsection{\label{sec:theory}The Theoretical Basis of FRD}
\begin{figure} [ht]
   \begin{center}
   \includegraphics[height=5cm]{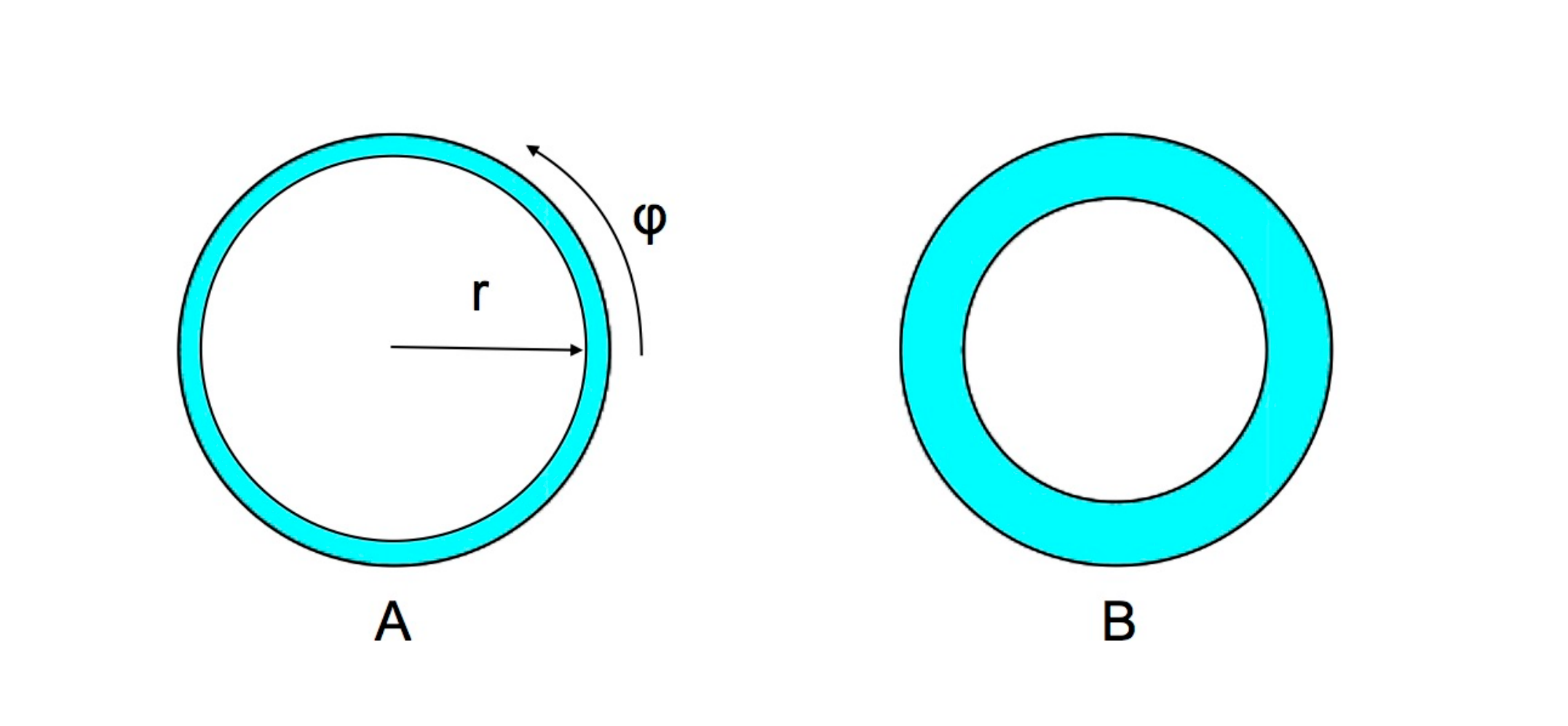}
   \end{center}
   \caption[goodfrd] 
   { \label{fig:goodfrd} 
Representations of the ring test results from two different fibres, using identical injection angles. Note that Fibre A has a much better FRD than Fibre B, where most of the rays are scattered azimuthally instead of radially, as in the case of Fibre B.}
\end{figure}

FRD is a result of fibre irregularities, or microbends\cite{car94}, and is heavily dependent on fibre termination procedures\cite{pop10}. Additionally, FRD has been shown to be wavelength dependent, though theoretical and experimental results disagree as to FRD performance in the blue and red regime\cite{bersh04,pop10,paz14}. FRD has the physical effect of decreasing the focal ratio of the input beam, resulting in beam spreading at the fibre output. This process can be described by the theoretical model of Gloge\cite{gloge72}, which uses the power flow through a fibre, modeled as a cylindrical waveguide, to examine the power loss within the fibre due to scattering at the core-cladding interface. Gloge also shows that by imaging the far-field (FF) intensity pattern, one can recover the modal power distribution of the fibre within which the associated FRD loss is encoded. Further work by Gambling, Payne \& Matsumura\cite{gamb75} and Carrasco \& Parry\cite{car94}, takes the model of Gloge and isolates a measurable quantity $D$ for the case of a well characterized input beam, in particular a collimated input beam. They also show that a the FF intensity pattern resulting from a collimated input beam is a ring, centered about the optical axis. The ring then has a finite thickness measurable as a Gaussian distribution. While the ring radius is related to the angle of the input beam, the thickness is related to the microbending measurable $D$. Two examples of representative FF intensity patterns are shown in Fig.\ref{fig:goodfrd}, with Fibre A showing a better FRD than Fibre B. Fig.\ref{fig:goodfrd} also shows that FRD is proportional to the number of rays scattered radially compared to those scattered azimuthally, with pure azimuthal scattering being ideal\cite{all13}. Therefore, by constructing an optical bench where a collimated beam is injected at various angles and for which the output FF intensity pattern is examined, FRD can be measured indirectly in a simple and consistent manner.  
\medskip
\subsection{\label{sec:ring}The Collimated Beam or ``Ring'' Test}
\begin{figure} [ht]
   \begin{center}
   \includegraphics[height=5cm]{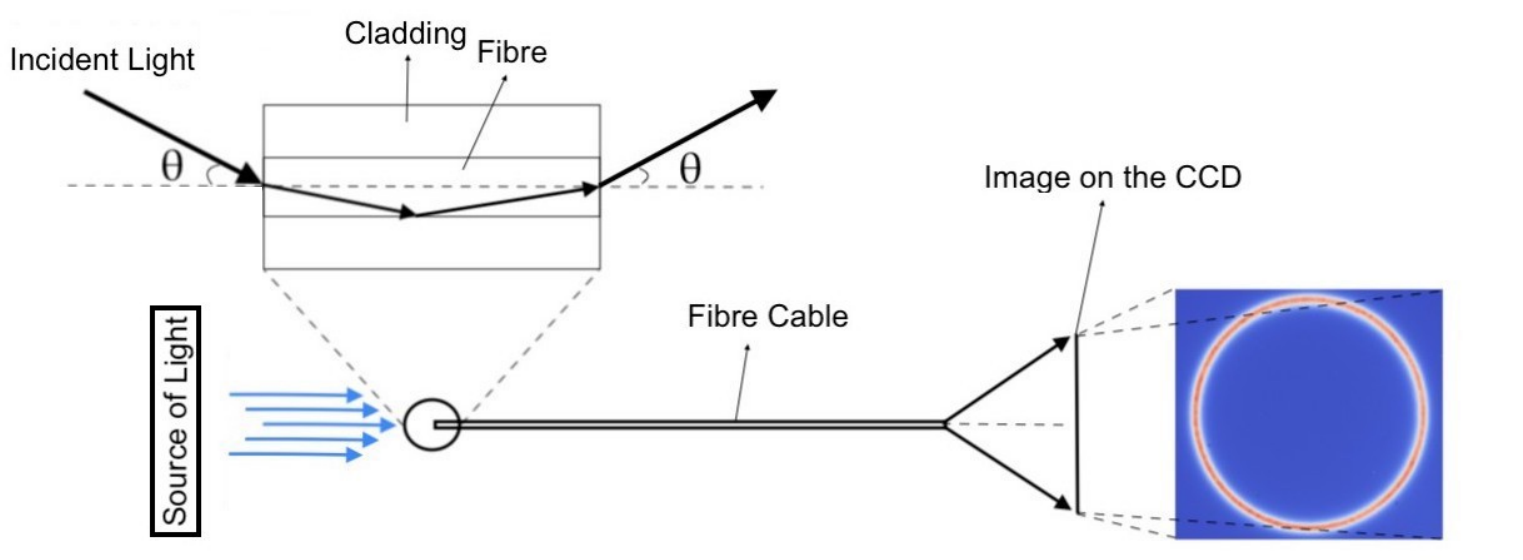}
   \end{center}
   \caption[simpring] 
   { \label{fig:simpring} 
Simplified schematic of the ring test showing the injection of the collimated beam at a given input angle and the resulting far-field ring pattern.}
\end{figure}

The method described in the above section forms the theoretical basis of the collimated beam or ``ring'' test. Fig.\ref{fig:simpring} is a simplified schematic of the ring test showing the injection of a collimated beam at a incident given angle and the resulting FF intensity pattern. Standard components of the ring test include; a single mode light source, intermediate fibre, collimator, input and output alignment stages, an imager and a fibre under test. By mounting the first alignment stage on a rotational stage, the injection angle incident on the test fibre can be varied. For each injection angle the FF ring is imaged, and the corresponding diameter and ring thickness Full Width Half Max (FWHM) are determined. The FWHM values are then converted to their corresponding 1/$e^2$\footnote{1/$e^2$ is when we have 13.5\% of the peak intensity} values. The ratio of this 1/$e^2$ value to the radius of the ring is then used to compute the relative FRD of the system. This method quantitatively determines the FRD for optical systems with different f/\#s. The FRD proxy measurement is defined by Eqn.\ref{eqn:frd}. Additionally, the injection f/\# may be determined for the system at each injection angle using Eqn.\ref{eqn:fin}. 

\begin{equation}\label{eqn:frd}
\text{FRD} = \frac{1/e^{2}}{D}
\end{equation}

\begin{equation}\label{eqn:fin}
f_{in} = \frac{1}{2\tan(\theta_{i})}
\end{equation}

\subsubsection{\label{sec:auto}The Case for Automation}
\begin{figure} [ht]
   \begin{center}
   \includegraphics[scale=0.51]{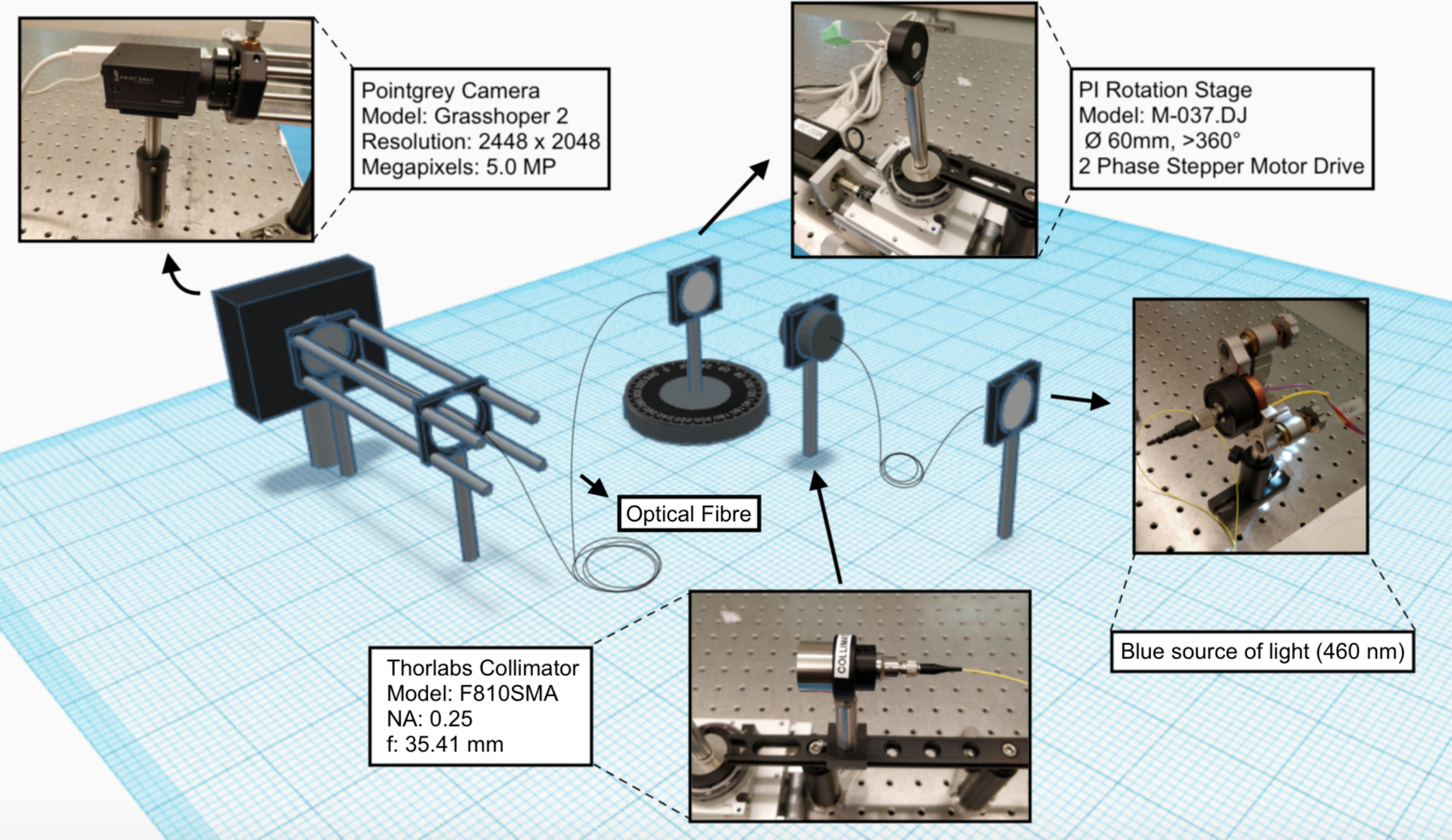}
   \end{center}
   \caption[fullschem] 
   { \label{fig:fullschem} 
Full schematic of the ring test optical bench components.}
\end{figure}

In the case of MSE FiTS, as the spectrograph is designed to sit at f/2, measuring a FRD proxy measurement for input angles around $\sim\theta_{i}=14^{o}$ is an effective method for ensuring the science requirements, namely FRD$<$5\%, are met for each of the $\sim4000$ fibres. Additionally, even with FRD$<$5\% at all times, if the FRD changes drastically during an observation of a single target, for example as a result of telescope elevation change or temperature, then the resulting change in the beam entering the spectrograph could reduce the precision of measurements such as radial velocity or spectrophotometry. This makes both the individual FRD measurements and the stability of the fibre FRD critical. A schematic of the full ring test set-up is shown in Fig.\ref{fig:fullschem}, where the components listed previously are labeled within. In general, to complete the ring test manually, one must: perform the required alignments to ensure even illumination of the fibre end, rotate the fibre end to the desired injection angle, adjust the camera characteristics visually to avoid saturation of the ring, capture the images using camera-specific software and finally, measure the diameter and thickness of the ring using an additional image analysis program. As this process must be repeated for a range of injection angles, the total time required to construct an FRD profile at various f/\#s can be substantial. Therefore, to test each MSE fibre individually and to avoid testing only a subset of the fibres, the process must be automated.

\section{\label{sec:bench}THE FITS FIBRE TEST BENCH}
The FiTS optical bench was constructed within the Mechanical Engineering Department at the University of Victoria under the supervision of faculty members and team members from NRC Herzberg Astronomy and Astrophysics Research Centre (NRC Herzberg). The bench was initially constructed to perform the ring test manually, testing fibres with known FRD characteristics to ensure repeatability with our bench. Components were then identified for automation in addition to identifying a consistent post-processing method and analysis procedure. The final optical bench was designed using existing hardware, free software available for download, is compatible with any operating system and is written entirely in Python. Here we present the complete system including; the automated hardware and methodology, the post-processing and analysis procedures and the master software wrapper.

\subsection{\label{sec:hard} Hardware Automation}
\begin{figure} [ht]
   \begin{center}
   \includegraphics[scale=0.5,angle=0]{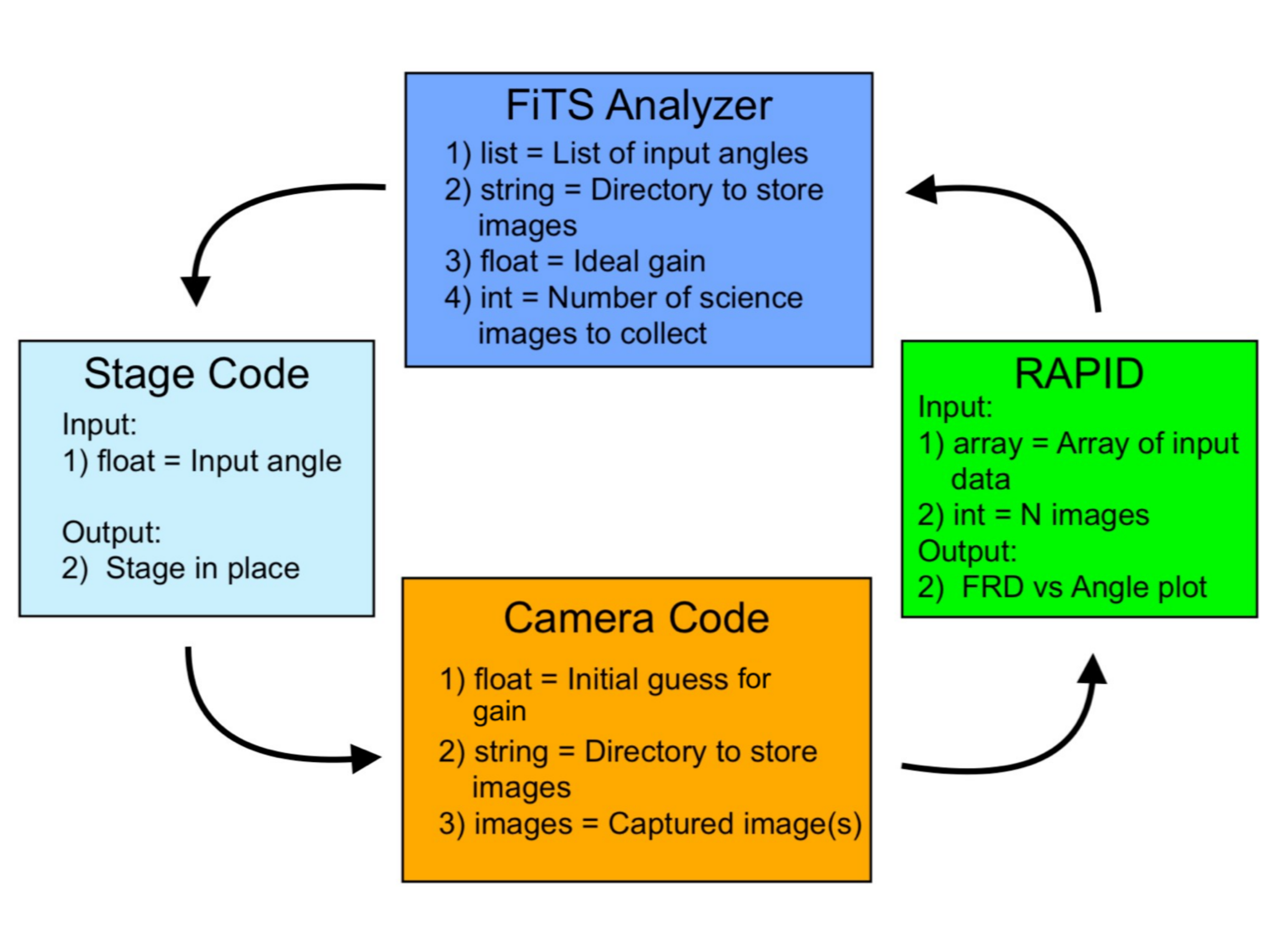}
   \end{center}
   \caption[flow] 
   { \label{fig:flow} 
Flowchart of the Big FFW with each box pertaining to a software module described in Sections \ref{sec:hard} and \ref{sec:soft}}
\end{figure}

Two hardware components were automated, the camera used for capturing the ring images and the rotational stage used to change the injection angle. Several cameras were available for use in the project but as the language of choice was Python, FLIR's Point Grey camera was selected for it's quality, free accompanying software FlyCapture and the companion Python wrapper PyCapture2. FlyCapture is a software development kit (SDK) for image capturing and camera control. It allows the user to select camera attributes such as the gain and exposure time (shutter speed), or allows the user to leave controlling the camera to the software itself. PyCapture2 can be imported into any Python program and be used to connect the camera, capture the images, format the images themselves and control camera characteristics. The stage was automated using a Mercury Controller by Physik Instrumente and accompanying Python wrapper PiPython. Using PiPython, the stage was connected, the current position was tracked and the movements were given in degree increments. Individual modules were created to control both the camera and the stage, with input and output shown in Fig.\ref{fig:flow}. 

\subsection{\label{sec:soft}Software Architecture: the Big FFW}
The modules for the camera, rotating stage and analysis are controlled by the master program we call the ``Big FiTS Fibre Wrapper'' (Big FFW)\footnote{The Big FFW will be available for download from the UVic FiTS team Github repository (\url{https://github.com/UVicOptics})}. The Big FFW is comprised of two layers. The first layer was written as a graphical user interface (GUI) using the built-in Python library TKinter, while the second layer manages the various automation modules and user input to produce the final results. A flowchart of the Big FFW is given in Fig.\ref{fig:flow}, with each of the automation modules shown as a block. The user input to the Big FFW is simply a list of incident angles, the camera gain, the number of images to take for each angle (in the case that the user would like to average to find the final results) and managerial paths to save the output images and final results.

Camera control was kept to a minimum, as we chose to use the auto exposure option of the FlyCapture. This decision was made after performing tests to examine the effect of different intensities on final FRD measurements. Images were taken using the auto exposure time option at various angles and found to differ in intensity by 62\% between initial and final injection angles, raising some initial concerns. Images were then collected after adjusting the exposure time by eye to match intensities between the initial and final injection angles, bringing the intensity difference down to 14\%. The resulting FRD measurements between the images taken using the auto exposure time option and taken by eye (with markedly different intensities), differed by $<1$\%. Therefore, we propose that changes in intensity in our system are not the dominant source of error in performing FRD measurements. Instead we attribute the dominant source of error in our system to non-homogeneities in ring illumination, which we have attempted to remove through accurate alignment and analysis routines. Additionally, the existence of a micro-lens array mounted on the face of the camera CCD introduced uneven illumination of one ring axis at large incident angles. This effect and possible solutions will be investigated further in the future but was neglected in this work as injection angles were kept below $14^{^\circ}$ equivalent to f/2 (see Eqn.\ref{eqn:frd}).

\subsubsection{\label{sec:rapid} Post-Processing Code: RAPID}
Analysis of the raw camera images output from the camera module was performed using our post-processing and analysis module the Ring Analyzer Python Interface for ring-test Data (RAPID). This interface chooses a thin horizontal slice from the output ring and examines the correlation between the pixel intensity and the pixel number of the data. As the raw data can suffer from significant background and systematic noise, RAPID smooths the data using a 1D Box filter kernel (convolution kernel). It does this by segmenting the data (with the number of segments depending on the amount of noise in the raw data), finding the mean intensity of the data in each segment and finally interpolating the data based on the determined mean. The non-isotropic nature of this filter makes it ideal for smoothing rings of different sizes. RAPID can also quantify the uniformity of the ring by computing the FRD of multiple slices of the ring and determining the standard deviation of the measurements.

\section{\label{sec:res}INITIAL RESULTS OF MSE-LIKE FIBRE CHARACTERIZATION}
\begin{figure} [ht]
   \begin{center}
   \includegraphics[scale=0.6]{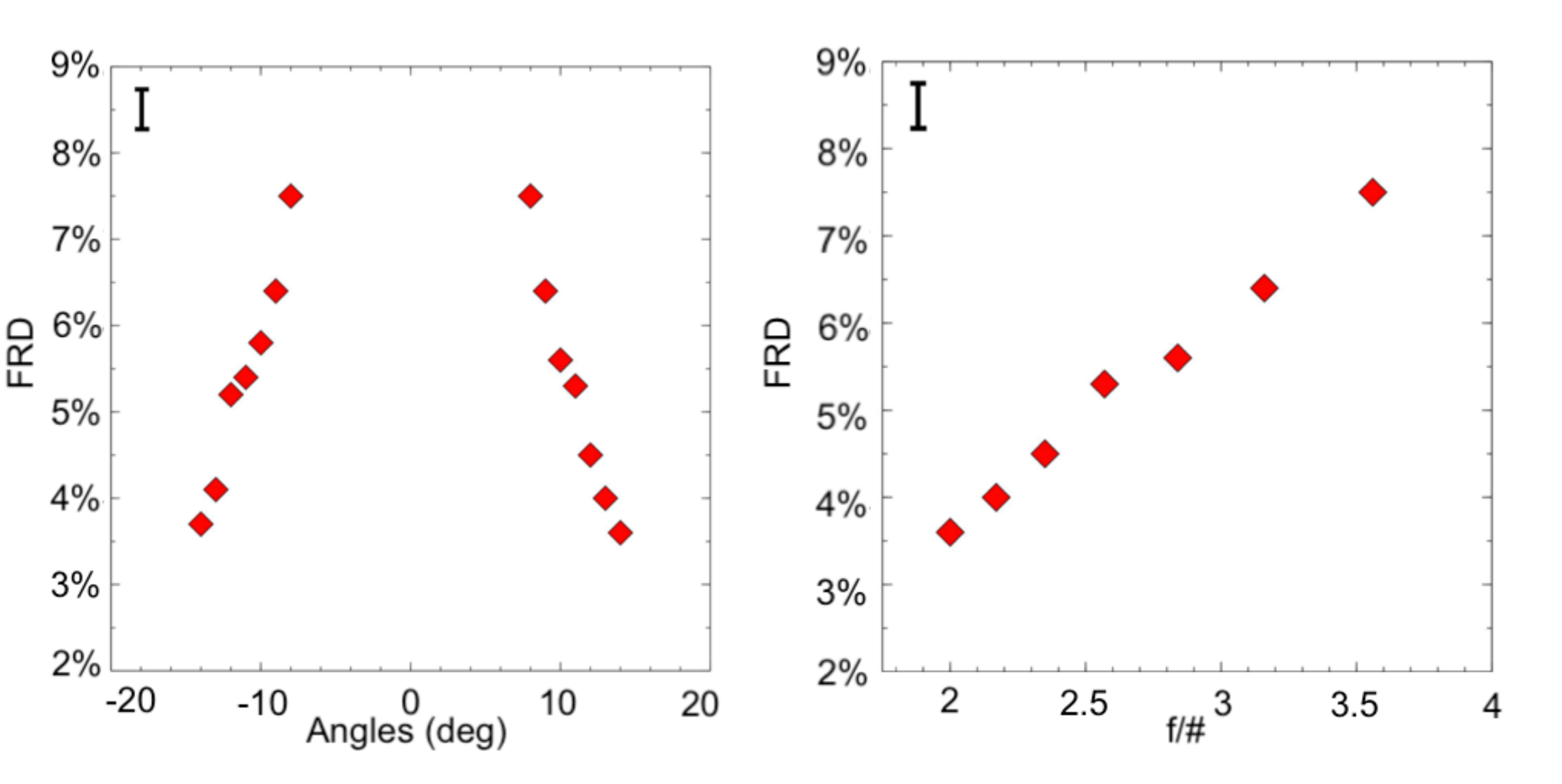}
   \end{center}
   \caption[flow] 
   { \label{fig:FRD_results} The FRD at 14 different incident angles found using the Big FFW (left panel). The symmetry in the plot suggests that our calibration process for finding the zero point has been successful. The f/\# plot of the positive angles are also calculated using Eqn.\ref{eqn:fin}. The right panel suggests that our estimated FRD of 3.7\% at f/2 is consistent with the science requirement of the MSE project. Error bars for all data points are pictured in the left upper corner of both panels.}
\end{figure}

\begin{figure} [ht]
   \begin{center}
   \includegraphics[scale=0.7]{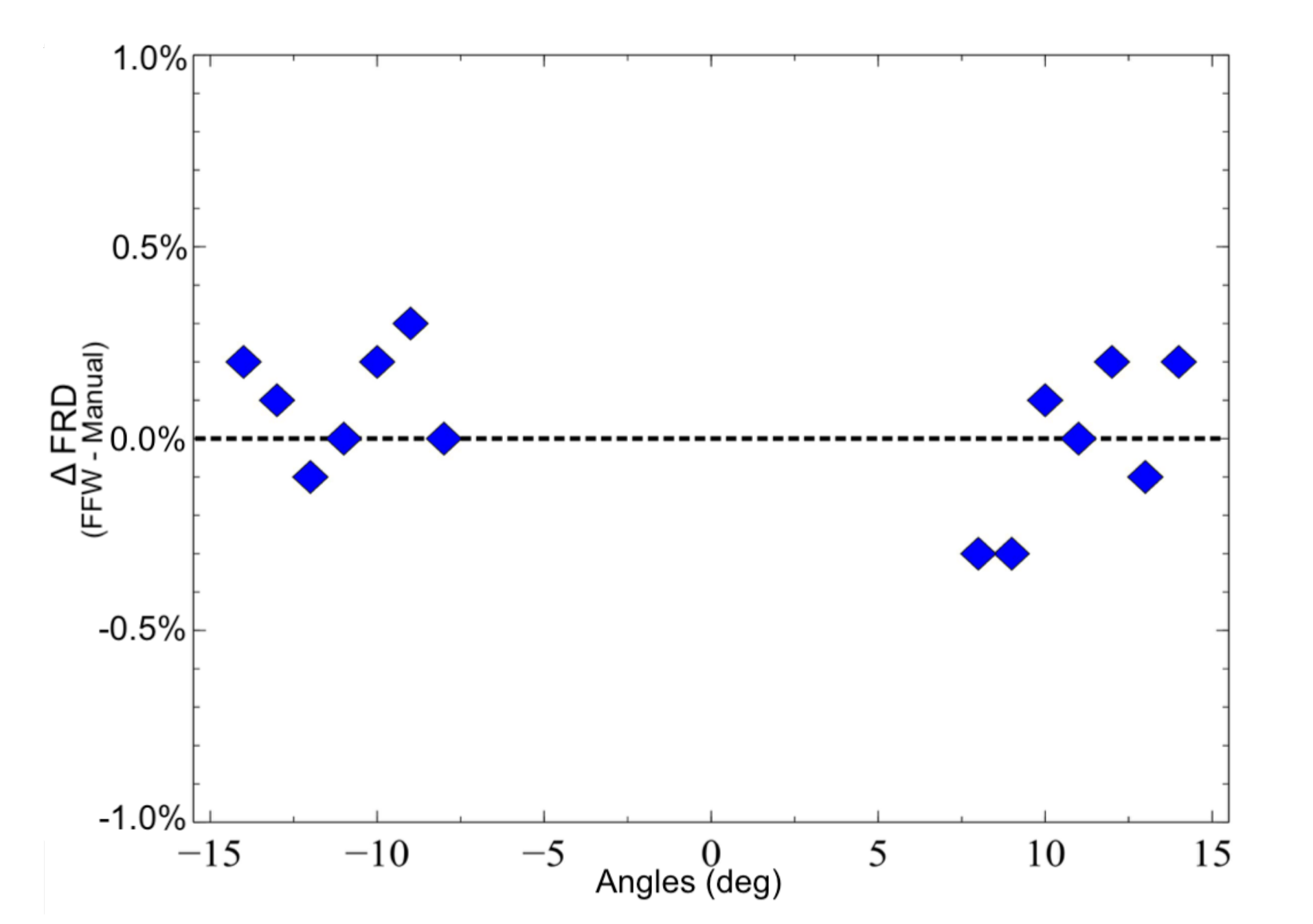}
   \end{center}
   \caption[flow] 
   { \label{fig:FRD_results2} The maximum difference of 0.3\% between the FRD values, suggests that the Big FFW’s results are in close agreement with results from individual/manual measurements of FRD.}
\end{figure}

A sample of the multi-mode GRACES fibre\footnote{provided by Fiber Tech Optica and NRC Herzberg} (see Pazder et al.\cite{paz14} for details on the fibre) was tested using the Big FFW on the FiTS optical bench. FRD-proxy measurements were made for 14 different incident angles (-14$^\circ$ to -8$^\circ$, and 8$^\circ$ to 14 $^\circ$), with the results shown in Fig.\ref{fig:FRD_results}. The current results suggest that we have met the science requirement of the MSE project (i.e. FRD of 3.7\% at f/2, which is well within 5\% science requirement of the MSE project). Additionally, the FRD of the MSE candidate GRACES fibre was determined manually and compared against the results determined by the Big FFW (see Fig.\ref{fig:FRD_results2}). We find the maximum disagreement between results from the Big FFW and the manual Ring Test to be less than the systematic error in our analysis (i.e. systematic error of 0.5\%).

\subsection{\label{sec:fut} Future Work}
Future work on the FiTS optical bench will involve upgrading the current design to facilitate the automated testing of both individual, and cabled fibres. Planned tests include quantifying fibre throughput, performing the ``formed beam'' or ``full aperture'' test and evaluating fibre FRD stability. These tests will be accomplished via modular additions to the existing system, with the ultimate goal of building a complete optical test bench for consistently testing the final 50m long MSE fibres. Preliminary work has already begun to develop a centroiding method to isolate individual fibres within a bundle. This will allow for testing of an entire bundle, while ensuring the fibre bundle remains intact. As work progresses, the Big FFW will remain the primary software used and will be made compatible for each test.

\subsubsection{\label{sec:futFA} Formed Beam Test}
\begin{figure} [ht]
   \begin{center}
   \includegraphics[scale=0.7]{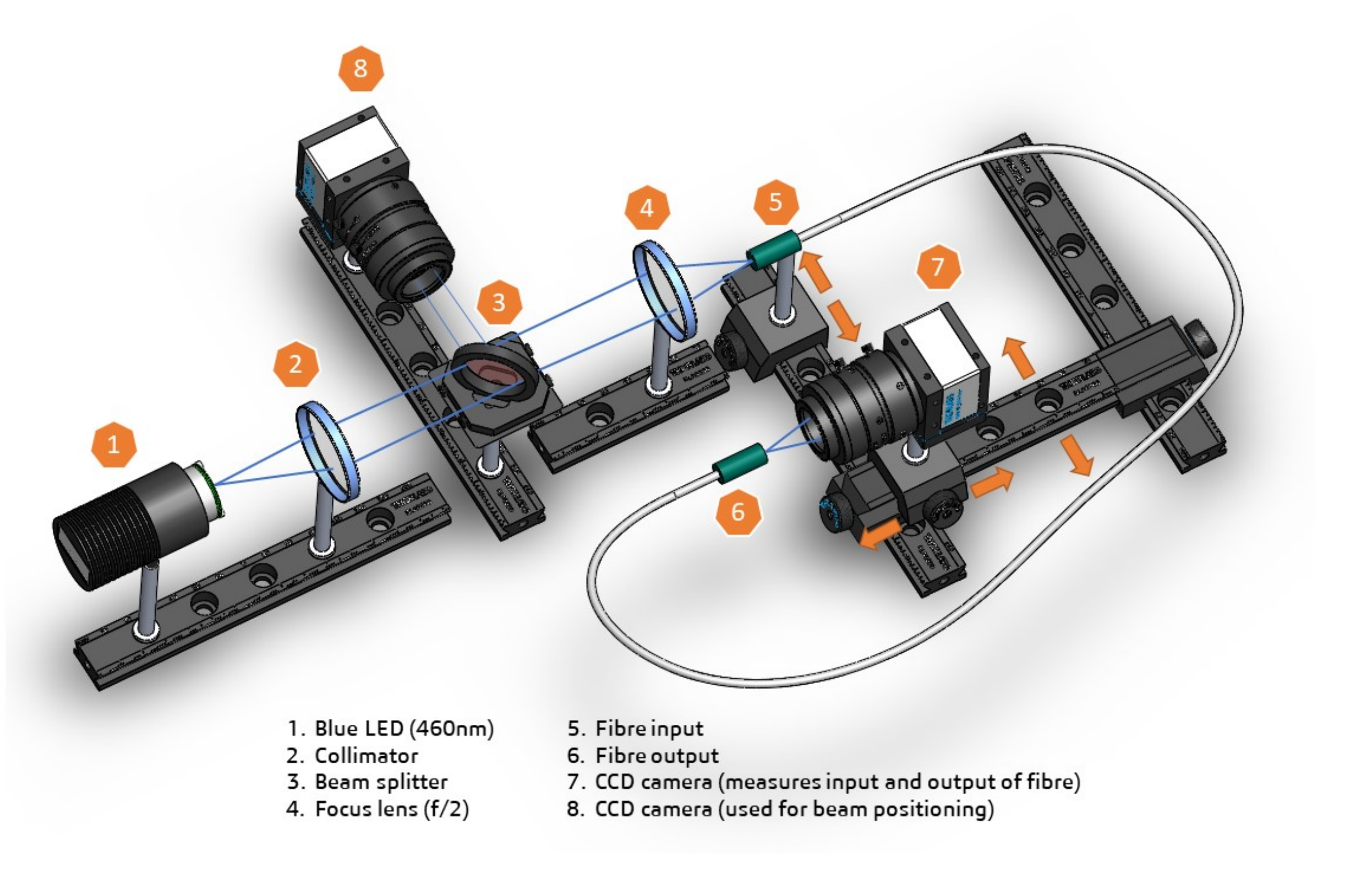}
   \end{center}
   \caption[] 
   { \label{fig:Throughput_system_concept}Formed beam and throughput tests concept design.}
\end{figure}

The ``formed beam'' test, to directly measure fibre FRD, will only be performed on fibre bundles and will closely resemble the set-up outlined by Pazder et al.\cite{paz14} and Carrasco \& Parry\cite{car94}. As described in Section \ref{sec:theory}, by injecting a beam at f/2 into the fibres and examining the output f/\#, a direct measurement of FRD may be made. A set-up similar to the one shown in Fig.\ref{fig:Throughput_system_concept} will be used to accurately position the focused beam at the input end of the fibre by means of a CCD camera (No. 8) allowing a visual inspection of the input end of the fibre (No. 5). Before each output measurement is made, another CCD camera (No. 7) will slide into view of the incoming beam to perform the input measurement. Immediately following, this camera will slide back into the position seen in the diagram to perform the output measurement, thus leaving room for the input end of the fibre to be accurately positioned to accept the incoming beam as previously described. The output f/\# will be determined by measuring the output ring diameter with respect to distance from the fibre output at 95\% encircled energy, as is standard. This could be accomplished by sliding the measurement camera (No.7) in Fig.\ref{fig:Throughput_system_concept} in the tangential direction to the fibre. By moving the camera closer and farther from the fibre output, the resulting output cone shape could be traced allowing us to investigate the possibility of light loss within the MSE optics. 

Finally, measurements of fibre throughput will also be made using a similar set-up to Fig.\ref{fig:Throughput_system_concept}. However, tangential motion of the measurement camera will no longer be necessary as a single measurement at the input and output ends of the fibre will provide the necessary information for this test.

\subsubsection{\label{sec:futSta} Stability}
FRD stability will be examined using fibre bundles and will involve subjecting the bundles to a variety of physical manipulations. This will be done to mimic the on-sky dynamics of the MSE as it moves through all its degrees of freedom. A sample of stability tests will include the following:
\begin{itemize}
  \item Ferrule tilting of $\pm2.2^{o}$ with a pivot radius of 250 millimeters
  \item Helical twisting of the bundle up to $\pm180^{o}$ over 800 millimeters
  \item Twisting of the bundle up to $90^{o}$ over 5 meters
  \item Rolling the bundles up to 1.3 meters in diameter
\end{itemize}

In addition to probing the FRD stability of the fibre bundles, these tests will also be used to investigate whether the demands of MSE on the bundles are reasonable and achievable. Finally, alternate light sources covering a broader wavelength regime will be introduced to examine FRD stability and dependency with wavelength. The ultimate goal will be to combine the throughput, formed beam and stability tests into a single automated system for complete and consistent fibre bundle testing. 

\section{\label{sec:sum}SUMMARY}
An optical test bench was designed, constructed and automated at the University of Victoria as part of the Maunakea Spectroscopic Explorer Fibre Transmission System (MSE-FiTS) project. The UVic FiTS optical bench performs the collimated beam, or "ring" test, on MSE candidate fibres to determine a proxy for the fibre FRD. 

In order to ensure the completion and consistency of future fibre tests of the more than 4000 MSE fibres, the FiTS optical bench was automated using existing (free) software and a specially written Python wrapper to control the various bench subsystems. Control of the camera and rotational stage was performed within the wrapper, the Big FFW, as well as the final post-processing and analysis. The final product is compatible with any operating system, entirely free and self-contained. The user is only required to enter a range of injection angles and one camera characteristic to the Big FFW. In return the FRD vs. incident angle profile is determined for the fibre under test. The Big FFW successfully reduces the time to test individual fibres from hours to minutes. If used by another team, the Big FFW can be adapted to communicate with any stage controller and camera.

Testing of a candidate MSE fibre was performed using the final FiTS optical bench and the Big FFW software. Tests performed using both the Big FFW and alternative software suggest that the candidate fibre meets the FRD requirements of MSE and verified that the Big FFW is a viable alternative to traditional Ring Test methods. 

\section{ACKNOWLEDGMENTS}
The authors and the MSE collaboration recognize and acknowledge the cultural importance of the summit of Maunakea to a broad cross section of the Native Hawaiian community.

SM, FJ, JL, KV and CB acknowledge the support of the NTCO CREATE program, NSERC funding reference number 498006-2017. PH acknowledges support from the Natural Sciences and Engineering Research Council of Canada (NSERC), funding reference number 2017-05983, and for sabbatical support from the National Research Council Canada.
\bibliography{report} 
\bibliographystyle{spiebib} 

\end{document}